\documentstyle[12pt,bezier]{article}
\def\odo{(\omega,\omega^{\dagger})}
\def\s2{{\hat {sl}}_2}
\def\j{j_{m.n}}
\def\bg{\beta\gamma}
\def\v{|0\rangle}
\def\cv{\langle 0|}
\def\f{{\cal F}}
\def\o{\omega^a(z)}
\def\oc{\omega_C^a(z)}
\def\doc{\omega^{\dagger}{}^b_{C^{\dagger}}(z)}
\def\zm{\sum_{m=1}^M}
\def\zn{\sum_{n=1}^N}
\def\tbb{\Upsilon_{a_1-1,b_m-1}({\bf z},{\bf u})}
\def\ibr{I_{{\bf a}{\bf b}}(\eta_2,...\eta_M,{\bf C})}
\def\1jc{\Upsilon_{{\bf a}{\bf b}}(z,C_1)}
\def\2jc{\Upsilon_{{\bf a}{\bf b}}(z,C_2)}
\def\cb{\Upsilon_{\bf a\bf b}(\bf z,\bf C)}
\def\pb{\Upsilon_{{\bf a},{\bf b}}(z,C)}
\def\pbk{\Upsilon_{\bf a\bf b}(z,C_k)}
\def\pbkk{\Upsilon_{\bf a\bf b}(z,C^k)}
\def\tcbt{\Upsilon_{{\bf a}{\bf b}}({\bf z},{\bf u})}
\def\tcbn{\Upsilon_{{\bf a}{\bf b}}({\bf z},{\bf u},{\bf C})}
\def\gnm{G_{a_n-1,b_m-1}({\bf z},{\bf u},{\bar {\bf z}},{\bar {\bf u}})}

\def\uf{f_{{\bf ab}}({\bf z},{\bf u})}
\def\tcb1{\Upsilon_{a_n-1,b_m-1}({\bf z},{\bf u})}
\def\pn{\prod_{n=1}^{N}}
\def\pm{\prod_{m=1}^{M}}
\def\on{\omega^{a_n}(z_n)}
\def\onz{\omega^{a_n}_{C}(z_n)}
\def\dom{\omega^{\dagger}{}^{b_m}(u_m)}
\def\domz{\omega^{\dagger}{}^{b_m}_{C_m^{\dagger}}(u_m)}
\def\dn{\frac{\partial}{\partial z_n}}
\def\dm{\frac{\partial}{\partial u_m}}
\def\gf{G_{{\bf ab}}({\bf z},{\bf u},{\bar {\bf z}},{\bar {\bf u}})}
\def\ozd#1{[\omega^{\dagger}(#1)\overline{\omega^{\dagger}(#1)}]}
\def\ozo#1{[\omega(#1)\overline{\omega(#1)}]}
\def\zzo#1{[\omega(#1)]}
\def\zzd#1{[\omega^{\dagger}(#1)]}

\def\NP#1#2{{\it Nucl.Phys.} {\bf B#1} (#2)}
\def\PL#1#2{{\it Phys.Lett.} {\bf B#1} (#2)}
\def\CMP#1#2{{\it Commun.Math.Phys.} {\bf #1} (#2)}
\vspace{2cm}

\title{{\bf Towards complex (rational) powers of free fields, generalized
$\bg$ systems and non-polynomial quantum field theory}}
\author {Oleg Andreev
\thanks{e-mail:andreev@ifh.de}\,\,
\thanks{Supported by DFG}
\\Humboldt-Universit\"{a}t zu Berlin,\thanks{On leave from Moscow Institute for
Physics and Technology}\\
Mathematisch-Naturwissenschaftliche Fakult\"{a}t I,\\
Institut f\"{u}r Physik, Invalidenstra\ss e 110,\\
 10099 Berlin, Germany}

\date{}
\begin{document}

\maketitle
\begin{abstract}
The $\bg$ system is generalized by complex(rational) powers of the fields,
which leads to a corresponding extension of the Fock space. Two different
approaches to compute the Green functions of the physical operators are
proposed. First the complex (rational) powers are defined via an integral
representation, that allows to compute the conformal blocks, Green functions
and
structure constants of OPA. Next an approach based on a system of
recursion equations for the Green functions is developed. A number of solutions
of the system is found. A lot of possible applications is briefly discussed.
\end{abstract}

\vspace{-15.5cm}
\hspace{12.5cm}
\begin{tabular}{ll}
HUB-IEP-94/9 \\
hep-th/9407xyz
\end{tabular}

\vspace{11.5cm}
\newpage
\section{Introduction}

In fact the problem addressed in this work is an old one. Quantum Mechanics
deals with Hamiltonians
$$
H=\Delta+U \qquad,
\eqno{(1.1)}
$$
where $\Delta$ is a kinetic term and $U$ is a potential. It should be noted
that
the choice of $U$ is rather unrestricted. On the other hand in Quantum Field
Theory $U$ is, as a rule, a polynomial of fields and their derivatives. The
problem is to extend a variety of Quantum Field Theory potentials to Quantum
Mechanical ones. It is clear that the extension will lead to new theories which
may give us a hope for realistic models of Particle Physics, Statistical
Mechanics etc.

In recent years, the most promising avenue to this extension has been to
explore
string field theory (see e.g.[1]). The Lagrangian of a string field $\Phi$ is
written in a form

$$
L=\langle\Phi|\hat Q|\Phi\rangle+\sum_{n=3}^{+\infty}\alpha_n\,\langle\Phi^n
\rangle \qquad,
\eqno{(1.2)}
$$
The first term is kinetic, the second represents a potential as an infinite sum
of string vertex functions. $\langle\Phi^n\rangle$ corresponds to a string
vertex function, such that the $n$ strings meet together.

There are a number of other examples. Some of them are provided by the 2d
Conformal Field Theory(CFT).

For illustration, let me consider the $SL(2)$ Wess-Zumino-Witten (WZW) model.
The well-known Wakimoto free field description of the model (at fixed point) is
built in terms of one free boson $\varphi$ coupled to a background charge and a
first order bosonic $\odo$ system of weight (0,1) [2], whose actions on the
plane are

$$
S_{\omega,\omega^{\dagger}}\sim\int d^2z\,\omega^{\dagger}\bar \partial\omega
\quad,\quad S_{\varphi}\sim\int d^2z\,\partial\varphi\bar\partial\varphi\qquad,
\eqno{(1.3)}
$$
where $\partial=\partial/\partial z$, ${\bar \partial}=\partial/
\partial{\bar z}$.
The currents are represented as

$$
J^{\dagger}(z)=\omega^{\dagger}(z)\qquad,
$$
$$
J^0(z)=-i(\omega\omega^{\dagger}(z)+\frac{1}{2\alpha_0}\partial\varphi(z))
\qquad,
\eqno{(1.4)}
$$
$$
J^-(z)=\omega^2\omega^{\dagger}(z)+ik\partial\omega(z)+\frac{1}{\alpha_0}
\omega\partial\varphi(z) \qquad,
$$
Here $k$ is the level, $2\alpha^2_0=1/(k+2)$.

The primary (physical) fields of the model are given by

$$
\Phi^j_j(z)=e^{-2ij\alpha_0\varphi(z)}
\eqno{(1.5)}
$$
for the highest weight vector
and

$$
\Phi^j_{-j}(z)=\omega^{2j}e^{-2ij\alpha_0\varphi(z)}
\eqno{(1.5a)}
$$
for the lowest weight vector.
\newline
$j$ is the weight of the representation. In the above I omitted
${\bar z}$-dependence.

{}From a physical point of view it is interesting to explore the weights $j$
which correspond to reducible (with singular vectors) representations of
the $\s2$ algebra. An irreducible representation is obtained by setting the
singular vectors to zero. This leads to differential equations for correlation
functions [3]. Kac and Kazhdan [4] found that there are singular vectors if
$j$ takes the values $\j$ defined by

$$
\j=\frac{m-1}{2}j_++\frac{n-1}{2}j_-\qquad,
\eqno{(1.6)}
$$
$$
\j=-\frac{m+1}{2}j_+-\frac{n}{2}j_-\qquad,
\eqno{(1.6a)}
$$
where $j_+=1,\,j_-=-k-2,\,\{m,n\}\in{\bf N},\,k\in{\bf C}$.

{}From this set it is worth to distinguish the so-called admissible
representations
[5], which correspond to the rational level $k$. In the case $j_-=-p/q$ it is
possible to recover the minimal models (series with $c<1$) via the
Drinfeld-Sokolov reduction. On the other hand $j_-=p/q$ leads to the Liouville
series with $c>25$. The second point is an existence of modular invariants for
such representations.

Now it is evident from (1.5a) and (1.6),(1.6a) that there are the physical
fields which are non-polynomial in the free fields! One has the free fields in
complex(rational, for the admissible reps.) powers.

Due to a connection of the $\s2$ algebra with $N=2$, parafermionic and
topological models (see e.g.[6] and refs. therein) it is expected that there
are
physical fields in such theories which are non-polynomial in free fields too.

For instance, a topological primary field in the Witten free field realization
[6] of the topological conformal algebra can be represented by

$$
\Psi_j=\varphi^{2j}\qquad,
\eqno{(1.7)}
$$
with $j$ is exactly given by (1.6). Note that the topological charge is given
by
\newline
$q=-2j/(k+2)$.

In this work I report a more handable problem, namely an extended $\odo$
system.
The extension is done by the complex(rational) powers of the free fields. This
model is of interest for several reasons. First of all, as it is known one can
use the model to build more complicated theories like $\s2 ({\hat {sl}}_n)$,
$N=2$, topological etc [6]. Second, it is a particular case of the so-called
$\bg$
systems [7] which play a crucial role in string theory. Of course, all results
presented may be generalized to an arbitrary $\bg$ system. The last reason is a
relative simplicity which allows one to focus on an effect of the
complex(rational) powers of the fields only.

The structure of the paper is as follows:

In section 2, the brief review of the $\odo$ system is given. In particular,
the
differential equations for correlation functions are represented. These
standard
equations follow from the $SL(2)$ invariance of vacua.

Section 3 provides a definition of the complex(rational) powers of the free
fields via an integral representation like the Mellin transform. By using this
definition a general conformal block of fields is computed. The Green functions
are defined through conformal blocks in the spirit of the 2d Conformal Field
Theory. As a result structure constants are calculated. Generalized Fock spaces
are proposed.

In section 4 alternative approach to the problem is developed. It is based on
other definitions for the fields which allow to obtain recursion equations for
the Green functions. A number of solutions is found and their correspondence
with the results of sec.3 is established.

The last, section 5, contains some conclusions and speculations.

\section{General properties of $\odo$ system}

As a preparation for a discussion of complex(rational) powers of the free
fields
in the later sections, let me briefly recall the main properties of $\odo$
system [7,8].

Consider the action

$$
S\sim\int d^2z\,\omega^{\dagger}\bar \partial\omega\,+\,(c.c)\qquad,
\eqno{(2.1)}
$$
where $\omega$ and $\omega^{\dagger}$ denote a pair of conjugate bosonic fields
of dimension $0$ and $1$ respectively. $(c.c)$ means a pair of complex
conjugate
fields $({\bar \omega},{\bar \omega}^{\dagger})$. This is a special case
$\lambda=1$ of the $\bg$ system with weights $(1-\lambda,\lambda)$.

The two-point function on the plane is normalized as \footnote{This
normalization is chosen to build the free field representation of $\s2$ [9].}

$$
\langle\omega(z)\omega^{\dagger}(z^{\prime})\rangle=\frac{i}{z-z^{\prime}}
\qquad.
\eqno{(2.2)}
$$

In terms of mode expansions one has

$$
\omega(z)=i\sum_{n=-\infty}^{+\infty} \frac{\omega_n}{z^n}\quad,\quad
\omega^{\dagger}(z)=\sum_{n=-\infty}^{+\infty} \frac{\omega_n^{\dagger}}
{z^{n+1}}\qquad.
\eqno{(2.3)}
$$

Canonical quantization gives the following commutation relations

$$
[\omega_n\,,\omega^{\dagger}_{-n}]=1\qquad.
\eqno{(2.4)}
$$

The stress tensor and central charge are

$$
T(z)=i\omega^{\dagger}\partial\omega(z)\qquad,
\eqno{(2.5)}
$$
$$
c=2\qquad.
\eqno{(2.6)}
$$

In terms of mode expansion $T(z)$ is given by

$$
T(z)=\sum_{n=-\infty}^{+\infty}\frac{L_n}{z^{n+2}}\quad,\quad L_n=\sum_
{m=-\infty}^{+\infty}m:\omega^{\dagger}_{n-m}\omega_m:\qquad.
\eqno{(2.7)}
$$

Define a vacuum $\v$ as

$$
\omega_{n+1}\v=\omega^{\dagger}_n\v=0\quad,\quad n\in\bf Z_+\qquad.
\eqno{(2.8)}
$$

It is easy to see that $\v$ is $SL(2)$ invariant. Indeed, by using (2.7) and
(2.8) one can check that

$$
L_1\v=L_0\v=L_{-1}\v=0\quad,\quad\v=|sl_2\rangle\qquad.
\eqno{(2.9)}
$$

Now let me introduce a conjugate vacuum $\cv$ as

$$
\cv\omega_{-n}=\cv\omega^{\dagger}_{-n-1}=0\quad,\quad\cv 0\rangle=1\quad,
\quad n\in\bf Z_+\qquad.
\eqno{(2.10)}
$$

It is found that $\cv$ is not $SL(2)$ invariant. Namely,

$$
\cv L_0=\cv L_{-1}=0\quad,\quad\cv L_1=\cv\omega_1\omega^{\dagger}_0\not=0
\qquad.
\eqno{(2.11)}
$$

It is known that the vacuum $\cv$ is expressed through the $SL(2)$ invariant
one
as

$$
\cv=\langle sl_2|\delta(\omega_0)\qquad,
\eqno{(2.12)}
$$
where $\delta(\omega_0)$ is the picture changing operator\footnote{I consider
the case when $\cv=\lim_{z_0\to\infty}\langle sl_2|\delta(\omega(z_0))$.}.

The correlation functions (conformal blocks) are defined as

$$
\tcbt=\cv\pn\on\pm\dom\v\qquad,
\eqno{(2.13)}
$$
where ${\bf a}=(a_1,...a_N),\, {\bf b}=(b_1,...b_M),\, {\bf z}=(z_1,...z_N),
\,{\bf u}=(u_1,...u_M), \{a_n,b_m\}\in{\bf Z}_+$.

The balance of charges (zero modes) is

$$
\zn a_n=\zm b_m\qquad.
\eqno{(2.14)}
$$

It should be stressed that the dimension of the conformal block space (2.13) is
given by $D=1$.

{}From the $SL(2)$ invariance of vacua one can derive the following equations

$$
(\zn \dn+\zm\dm)\,\tcbt=0\qquad,
\eqno{(2.15)}
$$
$$
(\zn z_n\dn+\zm u_m\dm+b_m)\,\tcbt=0\qquad,
\eqno{(2.16)}
$$
$$
(\zn z_n^2\dn+\zm u_m^2\dm+2b_m u_m)\,\tcbt+i\zn\zm\tcb1=0\qquad,
\eqno{(2.17)}
$$
where $\tcb1=\cv\omega^{a_1}(z_1)...\omega^{a_n-1}(z_n)...\omega^{\dagger}{}^
{b_m-1}(u_m)...\omega^{\dagger}{}^{b_M}(u_M)\v$.

Note that the last term in (2.17) is due to the picture changing operator.

The Fock space $\f$ is obtained by acting on the vacuum $\v$ with the mode
$\omega_0$ and all the negative frequency modes of the fields
$\omega,\omega^{\dagger}$\footnote{The $\odo$ system has infinitely many
non-equivalent Fock spaces. I choose one of them by defining the vacuum as in
(2.8). It is done in anticipation of an application to the $\s2$ algebra.}. The
basis of $\f$ is given by the states

$$
\omega_0^{A_0}\omega_{-1}^{A_1}...\omega^{\dagger}_{-1}{}^{B_1}\omega^
{\dagger}_{-2}{}^{B_2}...\v\quad,\quad\{A_n,B_m\}\in\bf Z_+\quad.
\eqno{(2.18)}
$$

To compute the physical Green functions one has to combine the conformal blocks
(2.13) with their conjugate ones as

$$
\gf=\tcbt\overline{\tcbt}\qquad.
\eqno{(2.19)}
$$

It is clear that so obtained Green functions are local i.e. they don't depend
on a mutual position of operators in the Euclidian region $\bar z=z^{\ast}$.
$\ast$ means complex conjugation.

\section{Generalized $\bg$ system. Complex(rational) powers of the fields}

The purpose of this section is to develop all the machinery of sec.2 in the
case of the complex(rational) powers of the fields. It is curious and
essential that it leads to nontrivial results.

Define the complex power of the fields as

$$
\o=\oint_{C}dt\,t^{-1-a}e^{t\omega(z)}\qquad,
\eqno{(3.1)}
$$
$$
\omega^{\dagger}{}^b(z)=\oint_{C}dt\,t^{-1-b}e^{t\omega^{\dagger}(z)}\quad,
\quad \{a,b\}\not\in{\bf Z}_-\qquad.
$$
The integration contours will be shown later. Also I suppressed normalization
factors. Note that the definition is, in fact, the Mellin transform [10]. It
was
used in [11] in order to define complex power of $SL(2)$ generators
(differential operators).

By using the two-point function (2.2) one can easily check that the so-defined
$\o$, $\omega^{\dagger}{}^b(z)$ are the primary fields with respect to the
stress tensor (2.6) with the conformal dimensions $0$ and $b$, respectively.

Now let me say a few more words about this definition. It is clear that $\o$,
$\omega^{\dagger}{}^b(z)$
are rather complicated objects and, in fact, depend on a pair variables
$(z,C)$.
In some sense it is like a construction introduced in [12] where the admissible
representation of $\s2$ is attached to a pair (point on a curve, Borel
subalgebra of the underlying finite dimensional Lie algebra which the
representation is induced from). I will denote them as $\oc$, $\doc$ below.

To clarify this formal definition, consider the correlator (conformal block,
see
(2.13))

$$
\tcbn=\cv\pn\onz\pm\domz\v\qquad,
\eqno{(3.2)}
$$
where ${\bf a}=(a_1,...a_N),\,{\bf b}=(b_1,...b_M),\,{\bf z}=(z_1,...z_N),\,
{\bf u}=(u_1,...u_M),\,\{a_n,b_m\}\not\in\bf Z_-$;
\newline
${\bf C}=(C_1,...C_N,C^{\dagger}_1,...C^{\dagger}_M)$. $\cv ,\v$ are vacua
defined in sec.2.

To get a non-zero result for the conformal block one has to take into
account the constraint

$$
\zn a_n=\zm b_m\qquad.
\eqno{(3.3)}
$$

It will be shown that only in this case the conformal block is non-zero. The
constraint (3.3) is a generalization of the balance of charges in correlator
$\omega\,,\omega^{\dagger}$ fields (see (2.14)).

{}From (3.1) one finds that

$$
\tcbn=\cv\pn\pm\oint_{C_n}dx_n\oint_{C_m^{\dagger}}dy_m\,x_n^{-1-a_n}
y_m^{-1-b_m}e^{x_n\omega(z_n)}e^{y_m\omega^{\dagger}(u_m)}\v\,\,.
\eqno{(3.4)}
$$

By using the two-point function of the $\omega\,,\omega^{\dagger}$ fields I
arrive at

$$
\tcbn=\pn\pm\oint_{C_n}dx_n\oint_{C_m^{\dagger}}dy_m\,
x_n^{-1-a_n}y_m^{-1-b_m}exp\{ix_ny_m/ (z_n-u_m)\}\qquad.
\eqno{(3.5)}
$$

Now one can use the definition (3.1) in order to integrate over
$y_m$ \footnote{I assume $N\leq M$. Otherwise one has to integrate over $x_n$
to
get the minimal number of integrals.}

$$
\tcbn=\pn\pm\oint_{C_n}dx_n\,x_n^{-1-a_n}\bigl(\sum_{k=1}^{N}x_k/(z_k-u_m)
\bigr)^{b_m}
\qquad.
\eqno{(3.6)}
$$

By changing the variables $x_1=x,\,x_2=xt_1,\,...x_N=xt_{N-1}$ I obtain

$$
\tcbn=\oint_{C_0}dx\,x^D\prod_{n=1}^{N-1}\pm\oint_{C_n}dt_nt_n^{-1-a_{n+1}}
\bigl(\frac{1}{z_1-u_m}+\sum_{k=1}^{N-1}\frac{t_k}{z_{k+1}-u_m}\bigr)^{b_m}
\qquad,
\eqno{(3.7)}
$$
where $D=\zm b_m-\zn a_n-1$.

Choosing $C_0$ around $0$ one gets\footnote{Assume $\zm b_m\geq\zn a_n$ and
take the limit $r\to 0$, here $r$ is a radius of an integration contour around
$0$. After this an analytic continuation is used.}

$$
\tcbn=\prod_{n=1}^{N-1}\pm\oint_{C_n} dt_nt_n^{-1-a_{n+1}}
\bigl(\frac{1}{z_1-u_m}+\sum_{k=1}^{N-1}\frac{t_k}{z_{k+1}-u_m}\bigr)^{b_m}
\qquad.
\eqno{(3.8)}
$$

It should be noted that the constrain (3.3) is crucial in order to get a
non-zero result. Because only in this case the integral over $x$ does not give
rise to zero.

Now let me turn to some simple examples.

$\underline{N=1}$. In this case the expression (3.8) reduces to

$$
\tcbt=\pm(z_1-u_m)^{-b_m}\qquad.
\eqno{(3.9)}
$$
The dependence on $\bf C$ is trivial. The dimension of a conformal block space
is given by $D=1$.

$\underline{N=2}$. The relevant correlator is

$$
\tcbn=\oint_{C}dtt^{-1-a_2}\pm\bigl(\frac{1}{z_1-u_m}
+\frac{t}{z_2-u_m}\bigr)^{b_m}\qquad,
\eqno{(3.10)}
$$

One can rewrite it as

$$
\tcbn=\bigl(\frac{z_1-u_1}{z_2-u_1}\bigr)^{a_2}\pm (z_1-u_m)^{-b_m}\ibr\qquad,
\eqno{(3.11)}
$$
where
$$
\ibr=\oint_{C}dv\,v^{-1-a_2}(1-v)^{b_1}\prod_{m=2}^{M}(1-\eta_m v)^{b_m}\qquad,
\eqno{(3.12)}
$$
with the harmonic ratios
$\eta_m=\frac{u_1-z_2}{u_1-z_1}\,\frac{u_m-z_1}{u_m-z_2}$.

The integral (3.12) looks as a conformal block of the minimal models with one
screening operator [13]. One can choose the integration contour in a similar
way. It is easy to see that the dimension of a conformal block space is given
by
$D=M$.

Let me consider the case $M=2$ in more detail. For convenience set
$u_1=\infty,\,
\newline
z_1=1,\,u_2=z,\,z_2=0$.
Then

$$
\langle \omega^{\dagger}{}^{b_1}_{C^{\dagger}_1}(\infty)\omega^{a_1}_{C_1}(1)
\omega^{\dagger}{}^{b_2}_{C^{\dagger}_2}(z)\omega^{a_2}_{C_2}(0)\rangle\sim
(1-z)^{-b_2}\oint_C dv\,v^{-1-a_2}(1-v)^{b_1}(1-\frac{z-1}{z}v)^{b_2}\,.
\eqno{(3.13)}
$$

To explore the behavior of the integral under $z\to 0$, take the basic
integration contours as shown in fig.1 (see [13])

\vspace{1cm}
\unitlength=1mm
\special{em:linewidth 0.4pt}
\linethickness{0.4pt}
\begin{picture}(111.67,36.33)
\put(50.00,30.00){\circle*{1.33}}
\put(70.00,30.00){\circle*{1.33}}
\put(90.00,30.00){\circle*{0.00}}
\put(90.00,30.00){\circle*{0.67}}
\put(90.00,30.00){\circle*{1.33}}
\bezier{96}(50.00,30.00)(60.00,36.33)(70.00,30.00)
\bezier{88}(90.00,30.00)(101.33,33.67)(111.67,33.00)
\put(59.33,33.00){\vector(1,0){1.00}}
\put(107.33,33.00){\vector(1,0){1.00}}
\put(51.33,25.33){\makebox(0,0)[cc]{$z/z-1$}}
\put(71.33,25.33){\makebox(0,0)[cc]{$0$}}
\put(91.00,25.33){\makebox(0,0)[cc]{$1$}}
\put(64.00,37.00){\makebox(0,0)[cc]{$C_1$}}
\put(111.00,37.00){\makebox(0,0)[cc]{$C_2$}}
\end{picture}
\vspace{-2cm}
\begin{center}
Fig.1:Contours used in the definition of the basic conformal blocks in the case
of $N=2,\,M=2$.  \end{center}

A simple algebra leads to

$$
\1jc\sim z^{-a_2}(1-z)^{a_2-b_2}F(-b_1,-a_2;1-a_2+b_2;z/z-1)\qquad,
\eqno{(3.14)}
$$
$$
\2jc\sim z^{-b_2}F(-b_2,-a_1;1+a_2-b_2;z/z-1)\qquad,
\eqno{(3.14a)}
$$
where $F$ is the hypergeometric function.
\newline
Under $z\to 0$ one has

$$
\1jc\sim z^{-a_2}(1+O(z))\qquad,
\eqno{(3.15)}
$$
$$
\2jc\sim z^{-b_2}(1+O(z))\qquad.
\eqno{(3.15a)}
$$

Because an arbitrary contour $C$ is a linear combination of $C_1$ and $C_2$,
$\pb$ is given by

$$
\pb\sim z^{-a_2}(1+O(z))+z^{-b_2}(1+O(z))\qquad.
\eqno{(3.16)}
$$
In above I omitted a relative factor.

It is evident that such asymptotic behavior corresponds to the following OP
Algebra

$$
\ozd{z}^{b_2}\ozo{0}^{a_2}=
$$
$$
=C_1(a_2,b_2)\ozd{0}^{b_2-a_2}/|z|^{2a_2}+...
+C_2(a_2,b_2)\ozo{0}^{a_2-b_2}/|z|^{2b_2}+...\qquad,
\eqno{(3.17)}
$$
where dots mean terms differ from the first by the integer powers of $z,\bar
z$.
$C_1$ and $C_2$ are the structure constants.

In the case of integer powers one term disappears, yielding the usual OPA.
\newline
For example

$$
\omega^{\dagger}{}^b(z)\omega^a(0)=\omega^{\dagger}{}^b(z)\oint_{C}
dt\,t^{-1-a}e^{t\omega(0)}\sim
$$
$$
\sim\frac{1}{z^b}\oint_{C}dt\,t^{-1-a+b}e^{t\omega(0)}+...\sim C_1(a,b)
\omega^{a-b}(0)/z^b+...\quad,\quad b\in{\bf N}\quad.
\eqno{(3.18)}
$$
In above I omitted the $\bar z$-dependence for the sake of simplicity.

Note that it is possible to choose the another basis for the conformal blocks
(3.13). In that case there is a diagonal monodromy under $z\to 1$.

$\underline{N=2,M=3}$. Choosing the contours as shown in fig.2

\vspace{1cm}
\unitlength=1mm
\special{em:linewidth 0.4pt}
\linethickness{0.4pt}
\begin{picture}(122.33,36.00)
\put(40.33,30.00){\circle*{1.33}}
\put(60.00,30.00){\circle*{0.00}}
\put(60.00,30.00){\circle*{1.33}}
\put(80.00,30.00){\circle*{1.33}}
\put(100.00,30.00){\circle*{1.33}}
\bezier{92}(60.33,29.67)(69.67,36.00)(80.00,29.67)
\bezier{92}(80.00,30.00)(89.67,36.00)(100.33,30.00)
\bezier{92}(100.00,30.00)(111.67,33.00)(122.33,33.00)
\put(113.33,32.67){\vector(1,0){1.00}}
\put(68.67,32.67){\vector(1,0){1.00}}
\put(88.67,33.17){\vector(1,0){1.00}}
\put(73.67,36.33){\makebox(0,0)[cc]{$C_1$}}
\put(108.67,36.33){\makebox(0,0)[cc]{$C_2$}}
\put(93.67,36.33){\makebox(0,0)[cc]{$C_3$}}
\put(42.33,26.00){\makebox(0,0)[cc]{$z/z-1$}}
\put(61.33,26.00){\makebox(0,0)[cc]{$z^{\prime}/z^{\prime}-1$}}
\put(81.33,26.00){\makebox(0,0)[cc]{$0$}}
\put(101.33,26.00){\makebox(0,0)[cc]{$1$}}
\end{picture}
\vspace{-2cm}
\begin{center}
Fig.2:Contours used in the definition of the basic conformal blocks in the case
of $N=2,\,M=3$.
\end{center}
one has
$$
\Upsilon_{{\bf ab}}(z,z^{\prime},C)=
\langle \omega^{\dagger}{}^{b_1}_{C^{\dagger}_1}(\infty)\omega^{a_1}_{C_1}(1)
\omega^{\dagger}{}^{b_2}_{C^{\dagger}_2}(z)
\omega^{\dagger}{}^{b_3}_{C^{\dagger}_3}(z^{\prime})\omega^{a_2}_{C_2}(0)
\rangle
\eqno{(3.19)}
$$
with
$$
\Upsilon_{{\bf ab}}(z,z^{\prime},C_1)\sim z^{\prime}{}^{-a_2}(1-z)^{-b_2}
(1-z^{\prime})^{a_2-b_3}
F_1(-a_2,-b_1,-b_2,1-a_2+b_3;\frac{z^{\prime}}{z^{\prime}-1},\frac{z-1}
{z^{\prime}-1}\frac{z^{\prime}}{z})\,\,,
\eqno{(3.20a)}
$$
$$
\Upsilon_{{\bf ab}}(z,z^{\prime},C_2)\sim
z^{-b_2}z^{\prime}{}^{-b_3}
F_1(-a_1,-b_3,-b_2,1-a_1+b_1;\frac{z^{\prime}}{z^{\prime}-1},\frac{z}{z-1})
\,\,,
\eqno{(3.20b)}
$$
$$
\Upsilon_{{\bf ab}}(z,z^{\prime},C_3)\sim
(1-z)^{-b_2}(1-z^{\prime})^{-b_3}
F_1(-a_2,-b_3,-b_2,1-a_2+b_1;\frac{z^{\prime}-1}{z^{\prime}},\frac{z-1}{z})
\,\,,
\eqno{(3.20c)}
$$
where $F_1$ is the hypergeometric function in two variable [14].

Note that in the above I have got the Picard's integral representation for
$F_1$
(single integral of Euler type) [14]. I fixed
$u_1=\infty,\,z_1=1,\,u_2=z,\,u_3=z^{\prime},\,z_2=0$.

For an arbitrary $M$ a basic conformal block is given by

$$
I_{{\bf ab}}(\eta_2,...\eta_M, C)\sim F_1(\alpha,\beta_2,...\beta_M,\gamma;
f_2,...f_M)\qquad,
\eqno{(3.21)}
$$
where $\alpha,\,\beta_i,\,\gamma$ are functions in $a_i,\,b_i$ and $f_i$ are
functions in $\eta_i$. $F_1$ is the hypergeometric function in $M-1$ variables.
I got its integral representation as a single integral of Euler's type.

The conformal blocks (3.8) for $N\geq 3$ are more involved. In particular, one
can
try to use an integral representation to turn them into more standard form
like the one introduced in [13].

In order to compute the physical Green functions one has to combine the
conformal blocks (3.8) with the ones from $\bar{\omega},\,
\bar{\omega}^{\dagger}$
fields (see (2.19)). This is now less trivial.
Define the Green functions as a monodromy invariant forms

$$
\gf=\sum_{C,\bar C}h_{C\bar C}\tcbn\overline{\tcbn}\qquad.
\eqno{(3.22)}
$$
Here $h_{C\,\bar C}$ is a metric on the space of conformal blocks. It is clear
that the so defined Green functions are local. Note that in the case of integer
powers the above definition becomes a simple product of analytic and
antianalytic factors (see (2.19)).

Now let me sketch a computation of the structure constants of the OPA (3.17).

The Green function to be calculated is\footnote{I shift parameters in order to
get rid of the degeneration.}

$$
G_{\bf a\bf b}(z,\bar z)=\langle
\ozd{\infty}^{a+\varepsilon}\ozo{1}^{b+\varepsilon}\ozd{z}^b\ozo{0}^a
\rangle \qquad,
\eqno{(3.23)}
$$
or
$$
G_{\bf a\bf b}(z,\bar z)=\sum_{C,\bar C}h_{C\bar C}\pb\overline{\pb}\qquad.
$$

The relevant conformal block is

$$
\pb=\oint_{C}dv\,v^{-1-a}(1-v)^{a+\varepsilon}\bigl(1/(1-z)+v/z\bigr)^b\qquad.
\eqno{(3.24)}
$$
Denote $I_k(z)=\pbk,\,I^k(z)=\pbkk$, where $k=1,2$ and fix the phases of
the integrands so that $I_k(z)\,,I^k(z)$ are real. The integration contours are
shown in fig.3.

\vspace{1.5cm}
\unitlength=1.00mm
\special{em:linewidth 0.4pt}
\linethickness{0.4pt}
\begin{picture}(111.33,36.00)
\put(50.00,30.00){\circle*{1.33}}
\put(70.00,30.00){\circle*{1.33}}
\put(90.00,30.00){\circle*{1.33}}
\bezier{92}(50.00,30.00)(60.00,36.00)(70.00,30.00)
\bezier{88}(90.00,30.00)(101.67,33.33)(111.33,33.33)
\bezier{92}(70.00,30.00)(80.00,35.67)(90.00,30.00)
\bezier{80}(50.00,30.00)(39.67,33.00)(30.00,33.00)
\put(36.00,32.67){\vector(1,0){1.00}}
\put(59.33,32.67){\vector(1,0){0.67}}
\put(79.67,32.67){\vector(1,0){0.67}}
\put(103.67,33.00){\vector(1,0){1.00}}
\put(60.00,36.00){\makebox(0,0)[cc]{$C_1$}}
\put(100.33,36.00){\makebox(0,0)[cc]{$C_2$}}
\put(77.33,36.00){\makebox(0,0)[cc]{$C^1$}}
\put(39.33,36.00){\makebox(0,0)[cc]{$C^2$}}
\put(51.33,27.00){\makebox(0,0)[cc]{$z/z-1$}}
\put(71.00,27.00){\makebox(0,0)[cc]{$0$}}
\put(91.00,27.00){\makebox(0,0)[cc]{$1$}}
\end{picture}
\vspace{-2cm}
\begin{center}
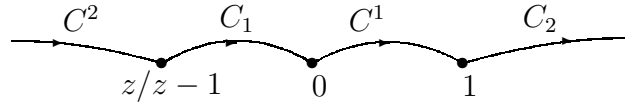
Fig.3:Two basic types of contours used for the definition of the conformal
blocks.
\end{center}
The basis $I_k$ corresponds to the diagonal monodromy of
the conformal blocks under $z\to e^{2i\pi}z$. The other has the diagonal
monodromy at $z=1$.

There is the following relation between them

$$
I_1=\frac{\sin \pi (a+\varepsilon)}{\sin\pi\varepsilon}I^1-\frac{\sin\pi
(b+\varepsilon)}{\sin\pi\varepsilon}I^2\qquad,
\eqno{(3.25)}
$$
$$
I_2=\frac{\sin\pi a}{\sin\pi\varepsilon}I^1-\frac{\sin\pi
b}{\sin\pi\varepsilon}I^2 \qquad.
\eqno{(3.25a)}
$$
The monodromy invariant Green function is

$$
G_{\bf a,\bf b}(z,\bar z)\sim\{I_1\bar I_1-\frac{\sin \pi(a+\varepsilon)}{\sin
\pi a}\frac{\sin \pi(b+\varepsilon)}{\sin\pi b}I_2\bar I_2\}\qquad.
\eqno{(3.26)}
$$

In terms of the normalized conformal blocks [3] it is rewritten as

$$
G_{\bf a\bf b}(z,\bar z)\sim \{B^2(-a,1+b)\f_1{\bar \f}_1
-NB^2(-b-\varepsilon,1+a+\varepsilon)
\f_2{\bar \f}_2\}\qquad.
\eqno{(3.27)}
$$
where $B$ means the B-function, $N=\frac{\sin\pi(a+\varepsilon)}{\sin\pi
a}\frac{\sin \pi(b+\varepsilon)}{\sin\pi b}$.

Finally normalizing the two-point function as

$$
\langle\ozo{z}^a\ozd{z^{\prime}}^a\rangle=1/|z-z^{\prime}|^{2a}\qquad,
\eqno{(3.28)}
$$
and taking the limit $\varepsilon\to 0$, one gets

$$
C_1(a,b)=C(a,b)\quad,\quad C_2(a,b)=C(b,a)\quad,\quad C(a,b)=
\frac{\Gamma(1+b)}{\Gamma(1+a)\Gamma(1-a+b)}\quad.
\eqno{(3.29)}
$$

Before ending this section, I wish to make several remarks.

First, in (3.1) complex powers are defined for $\{a,b\}\not\in{\bf Z}_-$. It is
not hard to see from (3.29) that fields in negative integer powers don't appear
in operator expansions.

In the case of $a(b)\in{\bf Z}_+$ there is a degeneracy i.e. it is
possible to build the Green function via $I_1(I_2)$ only. This leads to one
term
on the right hand side of (3.17) (also see (3.18)).

One can find the structure constants using analytic continuation from integer
$a,b$ to general ones. However in this case only one term appears on the
r.h.s. and the full OPA is disguised.

Finally, the Fock space $\f$ is extended to $\f^{(0)}$ by the complex(rational)
powers of $\omega,\,\omega^{\dagger}$. For this space the basis is given by the
states

$$
\omega_0^{A_0}\omega_{-1}^{A_1}...\omega^{\dagger}_{-1}{}^{B_1}\omega^
{\dagger}_{-2}{}^{B_2}...\v\qquad,
\eqno{(3.30)}
$$
where $\{A_0,\,B_1\}\in{\bf C}\backslash{\bf Z_-}$, the other powers are the
same as in (2.18).  Next one can define $\f^{(1)}$ with
$\{A_0,A_1,B_1,B_2\}\in{\bf C}\backslash{\bf Z}_-$.  It is implemented by the
complex(rational) powers of
$\omega,\,\omega^{\dagger},\,\partial\omega,\,\partial\omega^{\dagger}$ and so
on.

\section{Construction of conformal blocks (Green functions) via recursion
equations}

In the previous section, I have obtained the conformal blocks (Green functions)
of the $\omega,\,\omega^{\dagger}$ fields by the integral representation. I now
wish to develop an alternative approach to this problem, namely via a system of
recursion equations for the Green functions.

Consider the Green function

$$
\gf=\langle\pn\ozo{z_n}^{a_n}\pm\ozd{u_m}^{b_m}\rangle\qquad.
\eqno{(4.1)}
$$
For the sake of simplicity I will suppress the $\bar z,\bar u$-dependence for
the time being.

Define the complex(rational) powers of the fields as

$$
\zzo{z}^a=\omega(z)\zzo{z}^{a-1}\qquad,
$$
$$
\eqno{(4.2)}
$$
$$
\zzd{z}^b=\omega^{\dagger}(z)\zzd{z}^{b-1}\qquad.
$$

Next define the derivatives of these fields

$$
\frac{\partial}{\partial z}\zzo{z}^a=a\partial\omega(z)\zzo{z}^{a-1}\qquad,
$$
$$
\eqno{(4.3)}
$$
$$
\frac{\partial}{\partial z}\zzd{z}^b=b\partial\omega^{\dagger}(z)\zzd{z}^{b-1}
\qquad.
$$

Finally assume the following OP expansions

$$
\omega(z)\zzd{0}^b=ib\zzd{0}^{b-1}/z+O(1)\qquad,
$$
$$
\eqno{(4.4)}
$$
$$
\omega^{\dagger}(z)\zzo{0}^a=-ia\zzo{0}^{a-1}/z+O(1)\qquad.
$$
All definitions are ,of course, generalizations of the usual ones for the
arbitrary powers.

Now from (4.2) and (4.4) one can easily find

$$
\gf=\zn\frac{ia_n}{z_n-u_m}\gnm\qquad,
\eqno{(4.5)}
$$
$$
\gf=\zm\frac{ib_m}{z_n-u_m}\gnm\qquad,
\eqno{(4.5a)}
$$
where $\gnm=G_{a_1,...a_n-1,...a_N,\,b_1,...b_m-1,...b_M}
({\bf z},{\bf u},{\bar {\bf z}},{\bar {\bf u}})$.
\newline
It is clear that recursion equations in $\bar z,\bar u$ is obtained by a
similar way.

On the other hand (4.3) and (4.4) lead to

$$
\dm\gf=\zn\frac{ia_nb_m}{(z_n-u_m)^2}\gnm\qquad,
\eqno{(4.6)}
$$
$$
\dn\gf=-\zm\frac{ia_nb_m}{(z_n-u_m)^2}\gnm\,\,,
\eqno{(4.6a)}
$$
and the same equations in $\bar z,\bar u$.

Before solving the equations, let me point out some important consequences.

After a simple algebra with (4.5),(4.5a) the reader can derive

$$
(\zn a_n-\zm b_m)\gf=0\qquad.
\eqno{(4.7)}
$$
Using the above equation, one immediately deduces the constraint (2.13).

Algebraic manipulations with (4.6),(4.6a) give

$$
(\zn \dn+\zm\dm)\gf=0\qquad.
\eqno{(4.8)}
$$
By looking at this equation it is evident that it coincides with (2.15),
which reflects the $SL(2)$ invariance of vacua with respect to $L_{-1}$.

In addition to these equations, one may also derive the last two equations
following from the $SL(2)$ invariance (see (2.16),(2.17)). In order to do this,
one needs (4.5), (4.5a) as well as (4.6),(4.6a). After some algebra it is
possible to arrive at
$$
(\zn z_n\dn+\zm u_m\dm+b_m)\gf=0\qquad,
\eqno{(4.9)}
$$
$$
(\zn z_n^2\dn+\zm u_m^2\dm+2b_mu_m)\gf+i\zn\zm\gnm=0\,\,.
\eqno{(4.10)}
$$
It should be stressed that in (4.7)-(4.10) $\{a,b\}\in{\bf C}\backslash
{\bf Z_-}$.

A solution of the equations (4.5),(4.5a),(4.6),(4.6a) and corresponding
equations in $\bar z,\bar u$ is given by

$$
\gf=\sum_{ij}h_{ij}\Upsilon_{{\bf ab}}^i({\bf z},{\bf u})
\overline{\Upsilon_{{\bf ab}}^j({\bf z},{\bf u})}\qquad,
\eqno{(4.11)}
$$
where $\Upsilon_{{\bf ab}}^i({\bf z},{\bf u})$ is a solution of the system
(4.5)-(4.6a) and $\overline{\Upsilon_{{\bf ab}}^j({\bf z},{\bf u})}$ is
one of the system in $\bar z,\bar u$. $h_{ij}$ is a metric on the space of
solutions.

Now let me solve the system for the simple cases.

$\underline{N=1}$. Then (4.5)-(4.6a) are written in a form

$$
\tcbt=\frac{ia_1}{z_1-u_m} \tbb\qquad,
\eqno{(4.12)}
$$
$$
\tcbt=\zm\frac{ib_m}{z_1-u_m}\tbb
\eqno{(4.12a)}
$$
and
$$
\frac{\partial}{\partial z_1}\tcbt=-\zm\frac{ia_1b_m}{(z_1-u_m)^2}\tbb\qquad,
\eqno{(4.13)}
$$
$$
\dm\tcbt=\frac{ia_1b_m}{(z_1-u_m)^2}\tbb\qquad,
\eqno{(4.13a)}
$$
where $\tbb=\Upsilon_{a_1-1,b_1,...b_m-1,...b_M}({\bf z},{\bf u}).$
It is now a simple exercise to derive differential equations for the
conformal blocks $\cb$

$$
\dm\tcbt=\frac{b_m}{z_1-u_m}\tcbt\qquad,
\eqno{(4.14)}
$$
$$
\frac{\partial}{\partial z_1}\tcbt=-\zm\frac{b_m}{z_1-u_m}\tcbt\qquad.
\eqno{(4.14a)}
$$

The solution of eqs.(4.14),(4.14a) is given by

$$
\tcbt\sim\pm (z_1-u_m)^{-b_m}\qquad.
\eqno{(4.15)}
$$
It is the same as the one (3.9) derived via the integral representation.

$\underline {N=2, M=2}$.
After a substitution

$$
\tcbt=(z_1-u_1)^{b_2-a_1}(u_1-z_2)^{-a_2}(z_1-u_2)^{-b_2}\uf
$$

the equations (4.5)-(4.6a) are written in a form

$$
\uf=ia_1f_{a_1-1,b_1-1}({\bf z},{\bf u})-ia_2f_{a_2-1,b_1-1}({\bf z},
{\bf u})\qquad,
$$
$$
\uf=ia_1f_{a_1-1,b_2-1}({\bf z},{\bf u})-ia_2\frac{u_1-z_2}{u_1-z_1}
\frac{u_2-z_1} {u_2-z_2}f_{a_2-1,b_2-1}({\bf z},{\bf u})\quad,
$$
$$
\eqno{(4.17)}
$$
$$
\uf=ib_1f_{a_1-1,b_1-1}({\bf z},{\bf u})+ib_2f_{a_1-1,b_2-1}
({\bf z},{\bf u})\qquad,
$$
$$
\uf=-ib_1f_{a_2-1,b_1-1}({\bf z},{\bf u})-ib_2\frac{u_1-z_2}{u_1-z_1}
\frac{u_2-z_1} {u_2-z_2}f_{a_2-1,b_2-1}({\bf z},{\bf u})
$$

and

$$
\frac{\partial}{\partial u_1}\uf=\frac{z_{12}}{(z_1-u_1)(z_2-u_1)}
\{(b_1-a_2)\uf-ia_1b_1f_{a_1-1,b_1-1}({\bf z},{\bf u})\}\quad,
$$
$$
\frac{\partial}{\partial u_2}\uf=\frac{z_{12}}{(z_1-u_2)(z_2-u_2)}
\{(b_2-a_1)\uf+ia_1b_1f_{a_1-1,b_1-1}({\bf z},{\bf u})\}\quad,
$$
$$
\eqno{(4.17a)}
$$
$$
\frac{\partial}{\partial z_1}\uf=\frac{u_{12}}{(z_1-u_1)(z_1-u_2)}
\{(a_1-b_2)\uf-ia_1b_1f_{a_1-1,b_1-1}({\bf z},{\bf u})\}\quad,
$$
$$
\frac{\partial}{\partial z_2}\uf=\frac{u_{12}}{(z_2-u_1)(z_2-u_2)}
\{(a_2-b_1)\uf+ia_1b_1f_{a_1-1,b_1-1}({\bf z},{\bf u})\}\quad.
$$

A change of variables $(z_1,\,z_2,\,u_1,\,u_2)\to(z_1,\,z_2,\,\eta,\,u_2)$,
where $\eta=\frac{u_1-z_2}{u_1-z_1}\,\frac{u_2-z_1}{u_2-z_2}$
leads to

$$
\frac{\partial}{\partial z_1}\uf=\frac{\partial}{\partial z_2}\uf=
\frac{\partial}{\partial u_2}\uf=0\qquad.
\eqno{(4.18)}
$$

{}From this it follows that $\uf$ depends on $\eta$ only, i.e.
$\uf= f_{{\bf ab}}(\eta)\equiv f_{{\bf ab}}$. Now it is possible to
find equations

$$
\eta\frac{d}{d\eta}f_{{\bf ab}}=(a_2-b_1)f_{\bf a,\bf b}+ia_1b_1
f_{a_1-1,b_1-1}\qquad,
\eqno{(4.19a)}
$$
$$
\eta\frac{d}{d\eta}f_{{\bf ab}}=a_2f_{\bf a,\bf b}+ia_2b_1
f_{a_2-1,b_1-1}\qquad,
\eqno{(4.19b)}
$$
$$
\eta\frac{d}{d\eta}f_{{\bf ab}}=b_2f_{\bf a,\bf b}-ia_1b_2
f_{a_1-1,b_2-1}\qquad,
\eqno{(4.19c)}
$$
$$
\eta\frac{d}{d\eta}f_{{\bf ab}}=-ia_2b_2f_{a_2-1,b_2-1}\qquad.
\eqno{(4.19d)}
$$

Next combining (4.19a) with (4.19d) one gets

$$
\eta\frac{d^2}{d\eta^2}f_{{\bf ab}}+(1-a_2+b_1)\frac{d}{d\eta}
f_{{\bf ab}}-a_1a_2b_1b_2f_{{\bf a-1\,b-1}}=0\quad.
\eqno{(4.20)}
$$

The same procedure with (4.19b),(4.19c)  gives

$$
\eta^2\frac{d^2}{d\eta^2}f_{{\bf ab}}+(1-a_2-b_2)\eta\frac{d}{d\eta}
f_{{\bf ab}}+a_2b_2f_{{\bf ab}}-a_1a_2b_1b_2f_{{\bf a-1\,\bf b-1}}=0\quad.
\eqno{(4.20a)}
$$

Finally I obtain the Riemann equation

$$
\eta (1-\eta)\frac{d^2}{d\eta^2}f_{{\bf ab}}+\{(1-a_2+b_1)-(1-a_2-b_2)
\eta\}\frac{d}{d\eta}f_{{\bf ab}}-a_2b_2f_{{\bf ab}}=0\qquad.
\eqno{(4.21)}
$$

It is well known that the solution of this equation is given by [14]

$$
f_{{\bf ab}}=\oint_{C}dy\,y^{-1-a_2}(1-y)^{b_1}(1-\eta y)^{b_2}\qquad.
\eqno{(4.22)}
$$
It is not hard to see that the result coincides with the one obtained in sec.3.

Unfortunately computations for other cases are tedious. However it is expected
that they cover the results found in sec.3 via the integral representation.

\section{Conclusions and remarks}

First let me say a few words about the results.

At present the problem of extension a variety of Quantum Field Theory
potentials to Quantum Mechanical ones is an open problem. Some examples of
such extension are provided by the 2d Conformal Field Theory. Because
the free field representation plays a significant role in computations of the
conformal blocks, of the Green
functions and of Operator Algebras of the 2d Conformal Field Theory it seems
reasonable at first to extend the free field theories. Next one may build more
nontrivial theories via the free field representation by the standard methods
of CFT. The $\odo$ system is interesting for some reasons. First of all, it
has a
lot of applications in the 2d CFT, namely in $SL(N)$ WZW, $N=2$, parafermionic
and topological models [6]. Second this is a particular case of the $\bg$
systems
which play crucial role in String Theory. All the results for $\odo$ can be
generalized
to an arbitrary case. Finally it is a simple system that allows one to focus
purely
on the effect of non-integer powers.

In this work I have generalized the $\odo$ system by the complex(rational)
powers of the fields that leads to the corresponding extension on the Fock
space. Next two different approaches for a computation of the Green functions
are proposed. First the complex(rational) powers of the fields are defined via
the integral representation. It allows to compute the Green functions of the
physical operators in the spirit of the 2d CFT, namely through the conformal
blocks. The structure constants of OPA are calculated. Next the alternative
approach is developed. It is based on the system of recursion equations for the
Green functions. A number of solutions is found. They coincide with the
results obtained via the integral representation.

Let me conclude by mentioning some open problems.
\newline
(i) The main problem is, of course, to extend to more real theories, i.e.
models of the 4d Quantum Field Theory.

As to the 2d Field Theory one can, for example, introduce the following action

$$
S\sim\int d^2z\omega^{\dagger}\bar\partial\omega\,+\,(c.c)\,+\,g\int
d^2z[\omega\bar\omega]^a[\omega^{\dagger}\bar\omega^{\dagger}]^b\qquad,
\eqno{(5.1)}
$$
with a coupling constant $g$ and some noninteger parameters $a$,$b$. It is
clear that the results obtained in sec.3,4 allow to consider the perturbation
theory for such action.

In the context of topological models [6], one can define the action

$$
S\sim\int d^2z\omega^{\dagger}\bar\partial\omega\,+\,w^{\dagger}\bar\partial
w\,+\,(c.c)\,+\,g\int d^2z\,w^{\dagger}\bar
w^{\dagger}[\omega\bar\omega]^a\qquad,
\eqno{(5.2)}
$$
where $(w,w^{\dagger})$ is a first order fermionic system of weight $(0,1)$.
Now $a$ is a noninteger parameter that leads to the nonpolynomial potential
\newline
(ii) The second problem is impressive too. It concerns a lot of applications
of the techniques developed in the 2d CFT, for instance $\s2$, topological
models. Some
steps in this direction are available [15].
\newline
(iii) As it was shown in sec.3 there is no factorization for the operators
$\ozo{z}^a$,\newline$\ozd{z}^b$ in a general case. However it is clear that
they
decompose as

$$
\ozo{z}^a=\sum_{C,\bar C}\oc\overline{\oc}\qquad.
\eqno{(5.3)}
$$
It resembles a formula by Moore-Reshetikhin [16]. This time $C,\bar C$ play a
role of quantum group labels. At this point it would be interesting to
understand an underlying quantum group structure.
\newline
(iv) One puzzling aspect of the approach developed in sec.3 is that from the
definition (3.1) one has more contours than it is necessary to define the
conformal blocks\footnote{Indeed, as it was shown in sec.3 one can integrate
over ${\bf C} ({\bf C}^{\dagger})$ contours, as a result the conformal blocks
are defined in terms ${\bf C} ({\bf C}^{\dagger})$ only.}. This means that only
a part of them makes sense. To remedy this situation, one can bosonize the
$\odo$ system [7]. In terms of new variables ( scalar fields with background
charges) the $\omega,\,\omega^{\dagger}$ fields are written as

$$
\omega(z)=e^{\phi(z)+i\varphi(z)}\qquad,
\eqno{(5.4)}
$$
$$
\omega^{\dagger}(z)=\partial\varphi e^{-\phi(z)-i\varphi (z)}\qquad,
\eqno{(5.4a)}
$$

It is easy to check that the operator $\o$ corresponds to

$$
\o=e^{a(\phi(z)+i\varphi(z))}\qquad.
\eqno{(5.5)}
$$
Due to this the conformal blocks don't depend on $\bf C$. However this is not
the case for $\omega^{\dagger}{}^b(z)$.

It seems interesting to use bosonization in order to obtain integral identities
or something else in the spirit of the usual bosonization. It will be a
generalization of the last.
\newline
(v) Although I have considered the complex(rational) powers of the free fields
it is possible to develop such technique for more complicated objects. For
example, one can apply the approach of sec.4 to the Kac-Moody algebras.

\begin{center}
Acknowledgments
\end{center}
I am indebted from discussions with Vl.Dotsenko, B.Feigin, J-L.Gervais,
E.Martinec and C.Preitschopf. I thank G.Lopes Cardoso for reading the
manuscript. I would also like to acknowledge the hospitality of the LPTHE,
Paris $\rm {VI}$, where a portion of this work was completed. The research is
supported by DFG.


\begin{thebibliography}{99}

\bibitem{1}       M.Kaku, in Functional Integration, Geometry and Strings,
                  Birkhauser Press, Berlin (1989);{\it Phys. Rev.} {\bf D41}
                  (1990) 3734;\\
                  M.Kaku and J.Lykken, {\it Phys. Rev.} {\bf D38} (1988)
3052;\\
                  T.Kugo, H.Kunitomo and K.Suehiro, \PL{226}{1989} 48;\\
                  M.Saadi and B.Zwiebach, {\it Ann.Phys.} {\bf 192} (1989) 213.

\bibitem{2}       M.Wakimoto, \CMP{104}{1986} 604;\\
                  A.B.Zamolodchikov, talk given at Montreal (1988),
unpublished;\\
                  B.Feigin and E.Frenkel, {\it Lett.Math.Phys.} {\bf 19} (1990)
                  307; \CMP{128}{1990} 161;\\
                  D.Bernard and G.Felder, \CMP{127}{1990} 145.

\bibitem{3}       A.A.Belavin, A.M.Polyakov and A.B.Zamolodchikov, \NP{241}
                  {1984} 333.

\bibitem{4}       V.G.Kac and D.A.Kazhdan, {\it Adv.Math.} {\bf 34} (1979) 97.

\bibitem{5}       V.G.Kac and M.Wakimoto, {\it Proc.Natl.Acad.Sci.USA} {\bf 85}
                  (1988) 4956.

\bibitem{6}       A.B.Zamolodchikov and V.A.Fateev, {\it Sov.Phys.JETP}{\bf 62}
(1985) 215; {\it Sov.Phys.JETP} {\bf 63} (1986) 913;\\
                  C.Ahn, S.Chung and S.-H.Tye, \NP{365}{1991} 191;\\
                  E.Witten, On the Landau-Ginzburg Description of N=2 minimal
                  models, Preprint IASSNS-HEP-93/10;\\
                  A.M.Semikhatov, The MFF singular vectors in Topological
                  Conformal Theories, hep-th/9311180;\\
                  P.Fre, L.Girardello, A.Lerda and P.Soriani, \NP{387}{1992}
333.

\bibitem{7}       D.Friedan, E.Martinec and S.Shenker, \NP{271}{1986} 93.

\bibitem{8}       D.Friedan, {\it Notes on string theory and two dimensional
                  conformal field theory}, in Workshop on Unified String
Theories,
                  29 July-16 August 1985, eds. M.Green and D.Gross, World
                  Scientific (1986) 162;\\
                  E.Verlinde and H.Verlinde, {\it Lectures on String
Perturbation Theory},
                  delivered at the Trieste Spring School on Superstring, April
1988.

\bibitem{9}       Vl.S.Dotsenko, \NP{338}{1990} 747; \NP{358}{1991} 541.

\bibitem{10}      H.Bateman and A.Erdely, {\it Tables of Integral Transforms},
                  New York Toronto London McGraw-Hill Book Company, INC. 1954.

\bibitem{11}      B.Feigin and F.Malikov, Integral Intertwining Operators and
                  Complex Powers of Differential (q-difference) Operators,
                  Preprint RIMS-894 (1992).

\bibitem{12}      B.Feigin and F.Malikov, Fusion algebra at a rational level
and
                  cohomology of nilpotent subalgebras of $\s2$, hep-th/9310003.

\bibitem{13}      Vl.S.Dotsenko and V.A.Fateev, {\it Nucl.Phys.}
                  {\bf B240 [FS12]} (1984) 312; {\it Nucl. Phys.}
                  {\bf B251 [FS13]} (1985) 691.

\bibitem{14}      H.Bateman and A.Erdelyi, {\it Higher transcendental
                  functions}, New York Toronto London Mc Graw-Hill Book
Company,
                  INC.1953.

\bibitem{15}      O.Andreev, work in progress.

\bibitem{16}      G.Moore and N.Reshetikhin, \NP{328}{1989} 557.


\end{thebibliography}
\end{document}